\documentstyle[amssymb,aps,multicol]{revtex}

\newcommand{\cA}{{\cal A}}
\newcommand{\cC}{{\cal C}}
\newcommand{\cD}{{\cal D}}
\newcommand{\cE}{{\cal E}}
\newcommand{\cF}{{\cal F}}
\newcommand{\cG}{{\cal G}}
\newcommand{\cH}{{\cal H}}

\newcommand{\cL}{{\cal L}}

\newcommand{\cS}{{\cal S}}
\newcommand{\cZ}{{\cal Z}}

\newcommand{\cmplxs}{{\mathbb C}}

\begin{document}

\preprint{}
\draft
\title{Dynamical Generation of Noiseless Quantum Subsystems}
\author{Lorenza Viola${}^{1}$, Emanuel Knill${}^{2}$, 
and Seth Lloyd${}^{1\,\dagger}$ }
\address{ ${}^1$ d'Arbeloff Laboratory for Information Systems and 
Technology, 
Department of Mechanical Engineering,  \\
Massachusetts Institute of 
Technology, 
Cambridge, Massachusetts 02139 \\
${}^2$ Los Alamos National Laboratory, Los Alamos, New Mexico 87545} 

\maketitle

\begin{abstract}
We present control schemes for open quantum systems that combine
decoupling and universal control methods with coding procedures.  By
exploiting a general algebraic approach, we show how appropriate
encodings of quantum states result in obtaining universal control over
dynamically-generated noise-protected subsystems with limited control
resources.  In particular, we provide an efficient scheme for
performing universal encoded quantum computation in a wide class of
systems subjected to linear non-Markovian quantum noise and supporting
Heisenberg-type internal Hamiltonians.
\end{abstract}




\begin{multicols}{2}
Quantum bang-bang control has recently emerged as a general strategy
for manipulating quantum evolutions by enforcing suitable time scale
separations between the controller and the natural
dynamics of the system \cite{viola:1998}. For open quantum
systems, this has lead to establishing {\sl quantum error suppression}
schemes, whereby active decoupling from environmental noise is
achieved by continuously undoing system-bath correlations on time
scales that are short compared to the typical memory time of the bath
\cite{viola:1999a}.  Decoupling techniques were shown to be
consistent with efficient quantum information processing
\cite{viola:1999b}, thereby offering an alternative scenario compared
to error-correcting \cite{qecc} and error-avoiding quantum
codes \cite{eac}.  In contrast to the latter methods, {\sl no}
redundant encoding is necessary for preserving or manipulating quantum
information provided that the required control operations can be
implemented.  However, one may ask whether quantum coding
could be advantageous or necessary in situations where the available
control options are limited.

Answering the above question naturally connects the decoupling formalism with 
the notion of {\sl noiseless subsystem} that has been identified as 
the most general route to noise-free information storage \cite{knill:1999}. 
The basic philosophy is to envision the bang-bang control procedure as a 
tool for effectively endowing the system dynamics with a nontrivial group
of symmetries. Such symmetries generate structures in the system's state 
space which are in principle inaccessible to unwanted interactions 
and are therefore suited for encoding quantum information. 
Mathematically, the crucial requirement relates to the {\sl reducibility} 
properties of operator algebras associated with the action of the 
decoupling group. Variants of the same basic idea have been argued to lie 
at the heart of all existing approaches for stabilizing quantum 
information in a recent work by Zanardi \cite{zanardi:1999c}.

In this Letter we examine the implications of the above concept within
the decoupling framework, by showing that the action of the control
group allows for a complete classification of the choices available for 
both safe information encoding and universal control over coded states.  
At variance with the case where noiseless subsystems emerge by virtue of 
preexisting static symmetries in the overall Hamiltonian, the dynamical 
origin of the noise-protected structures also precisely constrains the 
admissible methods for implementing universal control in a way which 
simultaneously preserves the effect of decoupling as well as the selected 
coding space.

Using coding methods has several attractive consequences.  First,
bang-bang operations are needed only for noise suppression. Additional
manipulations on encoded subsystems become fully implementable via {\sl
weak strength} controls \cite{viola:1999b}. Second, for schemes where the
relevant Hamiltonians are allowed to be turned on or off slowly, an
advantage is that the corresponding pulses can be made more
easily frequency-selective.  Finally, coded states may be
intrinsically more robust against imperfections in the decoupler
operations. For a potentially large class of quantum information
processors characterized by {\sl linear} quantum noise, we outline a
scheme where noise-decoupling involves a minimal set of two
collective bang-bang rotations and universal quantum computation on
encoded qubits can be performed entirely through slow tuning of
two-body bilinear interactions.

{\it Decoupling.$-$} Let $S$ be a finite-dimensional quantum system
with self-Hamiltonian $H_S$ on $\cH_S$, dim$(\cH_S)=d$.
$S$ interacts with the environment $B$ via a Hamiltonian 
$H_{SB}=\sum_\alpha \, {E}_\alpha \otimes {B}_\alpha $,
the $B_\alpha$'s being linearly independent environment operators. 
The error operators ${E}_\alpha$ are assumed to belong to a 
linear space ${\cE}$ that we call the {\sl interaction space}. 
We require that tr($E_\alpha)=0$, thereby removing from $H_{SB}$ the 
internal evolution of the environment. Let $\cA_{\cE}$ denote the 
algebra generated by the identity, $H_S$, and $\cE$.
$\cA_{\cE}$ is a subalgebra of the full operator algebra
End$({\cal H}_S)$ closed under Hermitian transpose ($^\dagger$-closed). 
For $n$-qubit systems, $\cH_S \simeq {\cmplxs}^d$,
End$({\cal H}_S)\simeq \text{Mat}(d \times d, {\cmplxs})$, with $d=2^n$. 

In its essence, decoupling via bang-bang (b.b.) control relies on the 
idea of exploiting {\sl full strength/fast switching} control actions 
\cite{viola:1998,viola:1999a,viola:1999b}, meaning that a certain set of 
Hamiltonians can be (ideally) turned on/off instantaneously with arbitrarily 
large strength. Let $\cG$ denote a finite group determining the realizable
b.b. operations ({\sl decoupling group}), $|\cG|=\text{order}(\cG)$.
We identify the abstract group $\cG$ with its image under a unitary,
faithful representation $\mu$ by $d \times d$ matrices.  A decoupler
operates by subjecting the overall system to a cyclical time
evolution, the elementary temporal loop (of duration $T_c$ = {\sl
cycle time}) being designed to effect a suitable group-theoretical
averaging determined by $\cG$.  In the ideal limit of arbitrarily fast
cycle time, the action of the decoupler is equivalent to a modification of 
the effective dynamics according to $\cA_{\cE} \mapsto \Pi_{\cG}(\cA_{\cE})$, 
$\Pi_{\cG}$ being defined by \cite{viola:1999a,viola:1999b,zanardi:1999a}
\begin{equation}
X \mapsto \Pi_{\cG}(X) = {1 \over |{\cal G}| } 
\sum_{g \in {\cal G}} \, g^\dagger\,  X \, g \:,
\hspace{3mm} X \in \cA_{\cE}\:.
\label{projection}
\end{equation}
The quantum operation $\Pi_{\cG}$ is identical with the projector on the 
{\sl commutant} (or centralizer) of $\cG$ in End$(\cH_S)$, $Z(\cG)= 
\{{\cal O} \in \text{End}(\cH_S)\,|\,[ {\cal O}, g]=0 \;
\forall g \in {\cal G} \}$. Since $\Pi_{\cG}(H_S)=
H_{eff} \in Z(\cG)$, the decoupler essentially induces a 
$\cG$-symmetrization of the dynamics due to $H_S$. 

The commutant $Z(\cG)$ has a natural structure as a subalgebra of
End($\cH_S)$. A second algebraic structure associated with $\cG$ is
the algebra generated by $\cG$, ${\cmplxs}\cG$, which is the (at most)
$|\cG|$-dimensional vector space spanned by complex combinations of
elements in $\cG$ \cite{cornwell}.  Let ${\cmplxs}\cG'$ denote the
set of operators commuting with ${\cmplxs}\cG$.  Clearly,
${\cmplxs}\cG' = Z(\cG)$. The fact that both ${\cmplxs}\cG$ and
${\cmplxs}\cG'$ are $^\dagger$-closed subalgebras of End($\cH_S$) will
play an important role.  ${\cmplxs}\cG$ and ${\cmplxs}\cG'$ are linked
together by the property of {\sl reducibility} \cite{cornwell}.  
${\cmplxs}\cG$ is said to be {\sl
irreducible} (and $\cG$ to act irreducibly on $\cH_S$) if
${\cmplxs}\cG'=\{\lambda\cdot \openone \}={\cmplxs}\openone$.  A
similar definition applies to ${\cmplxs}\cG'$.  Since
${\cmplxs}\cG''={\cmplxs}\cG$, the non-triviality of $\cG$
automatically implies that ${\cmplxs}\cG'$ is reducible. Whether or
not ${\cmplxs}\cG$ acts irreducibly on $\cH_S$ distinguishes, at the
algebraic level, between maximal decoupling, where
${\cmplxs}\cG'={\cmplxs}\openone$, and selective decoupling, in which
case ${\cmplxs}\cG' \not = {\cmplxs}\openone$ \cite{viola:1999a}.
                                        
The goal of decoupling is to dynamically maintain evolutions of the system
so as to have a place where quantum information can safely reside and
undergo the required logical manipulations. The possibility to carry
out such a program without resorting to redundant encoding was
demonstrated in \cite{viola:1999b}. Is this the only relevant
situation?  

{\it Encoding.$-$} The basic idea is provided by the notion of a 
{\sl subsystem} \cite{knill:1999}. Mathematically, subsystems are 
identified as factors of subspaces by observing that the action 
of ${\cmplxs}\cG$ and ${\cmplxs}\cG'$ on $\cH_S$ can be represented as 
\cite{knill:1999,zanardi:1999c} 
\begin{eqnarray}
{\cmplxs}\cG & \simeq &\oplus_J\,
\openone_{n_J} \otimes \text{Mat}(d_J \times d_J, {\cmplxs})\:,
\label{gaction} \\  
{\cmplxs}\cG' & \simeq & \oplus_J\,
\text{Mat}(n_J \times n_J, {\cmplxs}) \otimes \openone_{d_J} \:,
\label{caction}  
\end{eqnarray}
where the index $J$ labels the $J$-th $d_J$-dimensional irreducible 
component of ${\cmplxs}\cG$, appearing with multiplicity $n_J$. Obviously, 
$\sum_J n_J d_J =d$. Such representations 
are associated with the following decomposition of $\cH_S$:
\begin{equation}
\cH_S \simeq \oplus_J \,\cH_J \simeq \oplus_J \,\cC_J \otimes \cD_J \:, 
\label{statespace}
\end{equation}  
with dim$(\cC_J)=n_J$, dim$(\cD_J)=d_J$. Results
(\ref{gaction})-(\ref{statespace}) stem from the general decomposition
theory of $^\dagger$-closed operator algebras. As argued in
\cite{knill:1999,zanardi:1999c}, they provide the common algebraic
ground for discussing noise control strategies.  In our setting, the
above relationships are linked to the decomposition of $\mu$ according
to the irreducible representations ({\sl irreps}) of $\cG$,
$\mu=\oplus_J n_J \mu_J $ \cite{cornwell}. Eq. (\ref{statespace})
reflects the fact that the subspace $\cH_J$ of states transforming
according to $\mu_J$ arises from $n_J$ replicas of a $d_J$-dimensional
irrep.  In a suitably chosen orthonormal basis of $\cH_J$,
$\{|J,l,m\rangle \,| \,l=1,\ldots, n_J; m=1,\ldots,d_J \}$, such a
one-to-one mapping is given by a correspondence of the form
$|l,m\rangle \leftrightarrow |l \rangle \otimes |m
\rangle$. Thus, the $J$-th eigenspace factorizes into the
tensor product of two factors $\cC_J$ and $\cD_J$, carrying irreps of
${\cmplxs}\cG'$ and ${\cmplxs}\cG$ respectively.
By construction, the dimensions of ${\cmplxs}\cG'$-irreps are found
as multiplicities of ${\cmplxs}\cG$-irreps, and vice versa.

The physical meaning behind the above construction is simple: 
The overall state space $\cH_S$ is decomposed into invariant
subspaces $\cH_J$, each of which can be regarded as the state space of
a bipartite system. For fixed $J$, $\cC_J$ is the state space of
a subsystem which is only acted on non-trivially by operators in
${\cmplxs}\cG'$, while $\cD_J$ is the state space of a subsystem which
is only acted on non-trivially by operators in ${\cmplxs}\cG$. Clearly, 
one is left with the freedom of exploiting {\sl any} of those subsystems for
encoding quantum states. Under what conditions is such an encoding noiseless ?

Let us first consider encoding in the left factors $\cC_J$
(``commutant coordinates''), assuming that $n_J>1$. When, in 
situations with underlying static symmetry, the decomposition 
(\ref{gaction}) is applied to the interaction algebra $\cA_{\cE}$, this 
generalizes the standard case of {\sl noiseless subspaces}, where coding 
takes place in the singlet sector of $\cA_{\cE}$, $d_{J_0}=1$, 
$\cC_{J_0} \simeq \cH_{J_0}$ \cite{eac,zanardi:1999c,zanardi:1999b}. 
Within the decoupling framework, protection against environmental noise 
is guaranteed if $\Pi_{\cG}(\cE)=0$ {\it i.e.}, ${\cal E}$ is correctable
by ${\cal G}$ \cite{viola:1999a,viola:1999b}. In fact, this condition 
is no longer necessary and can be replaced by the weaker 
requirement $\Pi_{\cG}(\cE) \in {\cmplxs}\cG' \cap {\cmplxs}\cG$, meaning 
that the effective error space belongs to the so-called {\sl center} of 
${\cmplxs}\cG$. Noise suppression is then ensured by the trivial action 
of the central elements on $\cC_J$, 
${\cmplxs}\cG' \cap {\cmplxs}\cG \simeq \oplus_J q_J \openone_{n_J} 
\otimes \openone_{d_J}$, $q_J \in {\cmplxs}$. Note that 
${\cmplxs}\cG' \cap {\cmplxs}\cG ={\cmplxs}\cG$ for Abelian decouplers. 

As a second coding method, we can choose the right factors $\cD_J$ 
(``group coordinates''). Such an option requires $d_J >1$, thereby 
excluding one-dimensional irreps. As a limiting case, this is the only 
possibility if $\cG$ acts irreducibly on $\cH_S$, in which case the 
decomposition (\ref{gaction}) collapses to a single term ${\cmplxs}\cG 
\simeq \text{Mat}(d \times d, {\cmplxs})$ and the whole space is a 
noiseless subsystem \cite{zanardi:1999c}. In general, since 
symmetrized noise generators $\Pi_{\cG}(E_\alpha) \in {\cmplxs}\cG'$ 
act trivially on factors carrying a ${\cmplxs}\cG$-irrep, subsystems 
of the form $\cD_J$ are automatically immune to environmental noise 
{\sl irrespective of the decoupler's ability to suppress the errors}. 
Although the overall effective dynamics is {\sl not} unitary in this case, 
corruption of states in $\cD_J$ is fully prevented due to their symmetry.

In addition to protecting against the environment, encoding may also
offer improved stability against faults in the implementations of b.b.
control. In particular, while imperfections of operations in $\cG$
directly affect the group component, states that carry 
${\cmplxs}\cG'$-coordinates are still unaffected as long as 
${\cmplxs}\cG'$ is preserved. Thus, encoding in the commutant degrees 
of freedom $\cC_J$ is {\sl robust} against imperfections of the b.b. 
rotations which stay in ${\cmplxs}\cG$. 
Experience from nuclear magnetic resonance suggests
that such imperfections do not severely affect the ability of the
decoupler to maintain noiselessness of the commutant degrees of
freedom. This effect will be analyzed elsewhere.

{\it Universal control.$-$} Since the group-theoretic averaging of the
decoupler is intrinsically associated with a minimum time scale $T_c$
\cite{viola:1999a,viola:1999b}, it is not surprising that control
operations are to be effected according to different timing criteria
depending on whether the intended action is on the group or the
commutant coordinates.  Regardless of the choice of $\cC_J$ or $\cD_J$
as the preferred coding space, transformations over a given subsystem
should not be allowed to ever draw states out of the protected
factor. This determines the symmetry of the Hamiltonians to be applied
for control, $H \in {\cmplxs}\cG'$ or $H \in {\cmplxs}\cG$ for action
on $\cC_J$- or $\cD_J$-subsystems respectively. Since the
application of Hamiltonians in ${\cmplxs}\cG'$ does not interfere with
the decoupler performances, encoding in $\cC_J$ has the virtue that
programming operations can be effected via the {\sl weak strength/slow
switching} scheme introduced in \cite{viola:1999b}. On the other hand,
when encoding in $\cD_J$ is chosen, slow application of arbitrary 
Hamiltonians produces a trivial action. The least demanding option for 
applying $H \in {\cmplxs}\cG - {\cmplxs}\cG'$ relies then on the 
ability of fast-modulating $H$ according to the {\sl weak strength/fast 
switching} scheme of \cite{viola:1999b}.

Let ${\cal U}(\cC_J)$ and ${\cal U}(\cD_J)$ denote the subgroups of
unitary transformations over the state space $\cC_J$ and $\cD_J$,
respectively. Universality results can be established by observing
that, by (\ref{gaction})-(\ref{caction}), ${\cmplxs}\cG'|_{\cC_J}
\simeq \text{Mat}(n_J \times n_J, {\cmplxs}) = \text{End}(\cC_J)$ and,
similarly, ${\cmplxs}\cG|_{\cD_J} \simeq \text{Mat}(d_J \times d_J,
{\cmplxs}) = \text{End}(\cD_J)$ {\it i.e.}, the elements of
${\cmplxs}\cG'$ (${\cmplxs}\cG$) restricted to the coding space span
the whole operator algebra of the associated subsystem. Thus, by
standard universality arguments \cite{lloyd:1995}, almost any pair of 
Hamiltonians $H_i \in {\cmplxs}\cG'$ or $H_i \in {\cmplxs}\cG$, $i=1,2$, 
is universal over $\cC_J$ or $\cD_J$, respectively. Similar 
existential results for control over commutant coordinates are 
formally derived in \cite{zanardi:1999b,zanardi:1999c}. 

If ${\cmplxs}\cG$ is irreducible, the possibility to attain complete 
control over $\cH_S$ \cite{viola:1999b} is directly found as a special 
case of the above results. When $\cG$ acts reducibly, reachability of 
arbitrary states in $\cH_S$ necessarily occurs through control operations 
that steer the system through different irreps of ${\cmplxs}\cG$ and 
${\cmplxs}\cG'$. The criteria for universality with no redundant encoding 
derived in \cite{viola:1999b} can then be regarded in terms of a symmetry 
mixing which arises from either combining commutant coordinates associated 
with different decouplers or from exploiting the action on both group 
and commutant coordinates of a single group $\cG$.
  
It is worth stressing that complete controllability of noiseless 
subsystems does not by itself imply the potential of efficiently 
implementing a quantum network. This depends on the available physical 
Hamiltonians as well as on the details of the architecture by which 
subsystems are actually configured to encode and process information.
We focus on quantum computation (QC).

{\it Universal quantum computation.$-$} Let $S$ be a quantum computer 
with $n$ qubits, ${\cal H}_S \simeq ({\cmplxs}^2)^{\otimes n}$. We 
consider henceforth a {\sl linear} interaction Hamiltonian of the form
\begin{equation}
H_{SB}= \sum_{a,i}\, \sigma_a^{(i)} \otimes B_a^{(i)}\;, 
\label{linear}
\end{equation}
for suitable environment operators $B_a^{(i)}$, $a=x,y,z$, $i=1,\ldots,n$. 
Eq. (\ref{linear}) encompasses various models of interest where the error 
space is spanned by single-qubit Pauli operators. Notably, the two extreme 
situations of independent and collective decoherence correspond to error 
generators of the form $\{E_\alpha\}=\{\sigma_a^{(i)} \}$, 
dim$({\cal E})=3n$, and $\{E_\alpha\}=\{\sum_i \sigma_a^{(i)} \}$, 
dim$({\cal E})=3$, respectively. 

{\sl Example 1: The collective spin-flips decoupling group}.  Let us
assume that $n$ is even and define $X_j=\sigma_x^{(j)}$,
$Z_j=\sigma_z^{(j)}$, with $Y_j=Z_j X_j= i \sigma_y^{(j)}$. The group
of collective $\pi$-rotations is the set $\cG =\{ \openone,
\otimes_{i=1}^n X_i,\otimes_{i=1}^n Y_i, \otimes_{i=1}^n Z_i \}$.
$\cG$ is an Abelian subgroup of the Pauli group for $n$ qubits, with
$k=2$ generators $\otimes_i X_i$, $\otimes_i Z_i$, $|\cG|=2^k=4$.  
Besides being identical with the stabilizer of distance-two $[n,n-2,2]$ 
error-correcting codes \cite{gottesman}, $\cG$ is also a subgroup
of the full group of collective rotations that plays the role of  
a generalized stabilizer for noiseless codes within the collective 
decoherence model \cite{bacon:1999}. Decoupling with $\cG$ is effective 
at suppressing any linear interaction of the form (\ref{linear}) 
since $\Pi_{\cG}(\sigma_a^{(i)})=0$.  
A single decoupling cycle is specified by a pulse sequence of the form 
$[\delta - {\cal P}_x - \delta - {\cal P}_z]^2$, $\delta =T_c/4$ and 
${\cal P}_a$ denoting a time delay and a collective $\pi$-pulse along 
the $\hat{a}$-axis respectively \cite{viola:1999b}. Since $\cG$ is Abelian, 
$\cG$ has $|\cG|=4$ one-dimensional irreps and the decomposition of 
$\cH_S$ is identical to the decomposition according to joint eigenspaces 
$\cH_J$, $J=1, \ldots, 2^k=4$, dim($\cH_J)=n_J=2^{n-k}$. 
Encoding into commutant factors $\cC_J$ is the only nontrivial option. 
Accordingly, each of the four (equivalent) joint $\cG$-eigenspaces is 
able to encode $n-2$ logical qubits.

Control operations over each $2^{n-2}$-dim noiseless subspace can be
implemented in the weak/slow fashion. Here is an explicit scheme for 
performing universal QC on encoded qubits. The key point is to look at the 
available operations in ${\cmplxs}\cG'$, which is easily done by exploiting 
the isomorphism of $\cG$ with the binary vector space $\cZ_2^{2n}$ along
with standard results from stabilizer theory \cite{gottesman}. As a
group, ${\cmplxs}\cG'$ has a set of $2n-2$ independent generators, two of 
which are also generators for $\cG$. The $2(n-2)$ generators of
${\cmplxs}\cG' - \cG$ can be chosen among interactions of the form $X_iX_j$,
$Z_iZ_j$, $i\not =j=1,\ldots,n$. These correspond to nontrivial encoded
operations.  For instance, the choice $\overline{X}_j=X_1 X_{j+1}$,
$\overline{Z}_j=Z_{j+1} Z_n$, $j=1, \ldots, n-2$, {\sl defines}
a set of $n-2$ logical qubits in terms of their encoded $\sigma_x$ and 
$\sigma_z$ observables \cite{note}.
A universal set of quantum gates is generated by observing that 
${\cmplxs}{\cG}'$ also contains the Heisenberg couplings 
$\vec{\sigma}_i \cdot \vec{\sigma}_j= X_iX_j + Y_iY_j +Z_iZ_j$
enabling one to implement swapping between any pair of {\sl encoded} qubits 
{\it i.e.}, $\vec{\sigma}_{\overline i} \cdot 
\vec{\sigma}_{\overline j}=\vec{\sigma}_{i+1} \cdot \vec{\sigma}_{j+1}$.
Since the square-root-of-swap gate together with one-qubit gates are 
a universal set \cite{divincenzo}, {\sl we can noise-tolerantly perform 
universal QC on $n-2$ encoded qubits by slowly turning on and off two-body 
interactions in parallel with the decoupler}.  

{\sl Example 2: The symmetric decoupling group}. Let $\cG =\cS_n$ be
the natural representation of the permutation group on the $n$-fold tensor 
product space $\cH_S$. In the presence of general linear interactions 
(\ref{linear}), decoupling according to $\cS_n$ forces effective permutation 
symmetry, thereby {\sl simulating} the collective decoherence model 
\cite{zanardi:1999c,zanardi:1999b,bacon:1999}.
Because $\Pi_{\cG}(\cE) = \{ \sum_i \sigma_a^{(i)}  \}$ 
$\subset {\cmplxs}{\cS}_n'$, 
noiseless subsystems are only supported by group factors $\cD_J$ carrying
${\cmplxs}{\cS}_n$-irreps. By recalling that ${\cmplxs}{\cS}_n'$ is 
identical with the algebra of totally symmetric operators generated by the 
global $su(2)$, the dimensions of such coding spaces can be calculated from 
the irrep multiplicities of angular momentum theory
\cite{cornwell,zanardi:1999c,zanardi:1999b}. Thus, 
dim($\cD_J)= (2J+1)n! / [(n/2+J+1)! (n/2-J)!]$, $J \in {\bf N}/2$.
An explicit scheme has been recently proposed for performing
universal QC on logical qubits encoded in clusters of $n=4, J=0$ 
physical qubits \cite{bacon:1999,note2}.
The same construction applies in our setting, with the additional
constraint that the exchange Hamiltonians required to implement 
universal gates should be fast-modulated at the same rate as the b.b. 
control within a cycle.

{\sl Example 3: The collective rotations decoupling group}. Let $\cG$ be 
the continuous group generated by the Lie algebra ${\cal L}=su(2)$ of
collective spin operators. Decoupling according to $\cG$ can be achieved by
performing the quantum operation (\ref{projection}) with respect to a suitable
finite-order symmetrizing group of unitaries ${\cal F}$, whose explicit 
form is given in \cite{zanardi:1999b}. 
Since $Z(\cG)=Z(\cL)={\cmplxs}\cS_n$, ${\cmplxs}\cS_n$-irreps emerge here
as commutant factors, making this example the dual of the previous one.
However, being $\Pi_{\cG}(\cE)=\Pi_{\cF}(\cE)=0$, noiseless subsystems
can be supported now by both commutant factors, in which case 
dim($\cC_J)=(2J+1)n! / [(n/2+J+1)! (n/2-J)!]$, or by group factors, for 
which dim($\cD_J)=2J+1$. In particular, if a $J=0$ four-qubits encoding 
in $\cC_J$ is chosen as above, the scheme for universal QC proposed by 
\cite{bacon:1999} can be fully implemented according to weak/slow 
control.

{\it Discussion.$-$} We presented dynamical procedures for generating and
controlling sectors of the state space of a generic open quantum
system, which are (ideally) immune to environmental noise. In addition 
to substantially expanding the range of possibilities for using active
decoupling methods, our analysis sheds light on the connections with
passive error protection schemes, where the relevant degrees of
freedom are decoupled from the noise-inducing interactions by virtue
of preexisting symmetries.  The presence of nontrivial symmetries is
found to be at the root of both active and passive stabilization
methods, thereby enabling the identification of common algebraic
structures. In spite of the mathematical resemblance, however, the two
strategies are physically very different. In particular, the limit of
long reservoir correlation {\sl length}, which underlies passive error
prevention in the presence of collective noise \cite{eac}, is replaced
by the dynamical requirement of long reservoir correlation {\sl time}
in active decoupling, which explicitly relies on the non-Markovian
nature of quantum noise \cite{viola:1999a}. The combination of 
decoupling and coding procedures results in a scheme for performing
universal quantum computation on noise-protected subsystems which is
highly appealing in terms of both the attainable encoding efficiency
and the overall control resources. Even in the limit where environmental
noise is fully tolerated, the scheme is not guaranteed to be robust
against arbitrary errors due to imperfect control. The performance of
decoupling in the presence of faulty control implementations along with 
the stability properties of the corresponding dynamically generated
subsystems will be discussed in a forthcoming work.

L. V. is grateful to D. P. DiVincenzo for inspiring discussions on 
stabilizer codes. This work was supported by DARPA/ARO under the QUIC 
initiative. E. K. received support from the DOE, under contract 
W-7405-ENG-36, and from the NSA.

${}^\dagger$ vlorenza@mit.edu; knill@lanl.gov; slloyd@mit.edu

\end{multicols}

\begin{thebibliography}{99}
\bibitem{note}
In terms of basis states, the code subspace is spanned by the $\cG$-invariant
$n$-qubit cat-states $(|x \rangle + | \textsc{not} \,x \rangle)/\sqrt{2}$, 
$x$ denoting an even-weigth binary string of length $n$.
\end{thebibliography}
\end{document}